# Recent CP violation and |V$_{us}$| measurements from the NA48 experiment

Cristina Biino  (for the NA48 collaboration)
*INFN, via P. Giuria 1, 10125 Torino, ITALY*

Over the last decade the NA48 collaboration at CERN has carried out an extensive experimental program dedicated to study CP violation (CPV) and rare processes in both neutral and charged kaon decays. A selection of results from recent analysis is compared with other measurements and theoretical predictions to provide constraints on CPV and on the CKM matrix. The observable η+− is related to the parameters of indirect and direct CPV (η+− = ε +  ε') and is defined as the CP violating amplitude ratio of the neutral kaon decaying into two charged pions, η+− = A(K$_L$→π$^+$π$^-$) / A(K$_S$→π$^+$π$^-$). NA48 has determined | η+− | by measuring the ratio of decay rates Γ(K$_L$→π$^+$π$^-$)/Γ(K$_L$→πeν). We obtain | η+− | = (2.223 ± 0.012)·10$^{-3}$.

The main objective of NA48/2 has been to search for direct CPV by high precision measurement of the asymmetry A$_g$= (g$^+$ - g$^-$) / (g$^+$ + g$^-$) of the linear slope parameter g in the Dalitz plot of K$^±$→3π  decays. NA48/2 used simultaneous K$^+$/K$^-$ beams which allow a high level of control of systematic effects. From the large data sample (3 ·10$^9$ K$^±$→π$^±$π$^+$π$^-$ and 9 ·10$^7$ K$^±$→π$^±$π$^0$π$^0$) we obtained the final results A$^c_g$ = (-1.5 ± 2.1) 10$^{-4}$ and A$^0_g$ = (1.8 ± 1.8) 10$^{-4}$ compatible with SM predictions. These results correspond to more than one order of magnitude improvement in precision with respect to previous measurements and are limited by the statistics of the data sample.

The decay rates R(Ke3/K2π), R(Kμ3/K2π) and R(Kμ3/Ke3) have been measured with charged K decays from special data samples with a low bias trigger in the NA48/2 experiment. Using the latest result for the K$^±$→π$^±$π$^0$ normalization channel, the Ke3 and Kμ3 branching fractions are then used as input to deduce the CKM matrix element V$_{us}$.

## 1.  EXPERIMENTAL SET-UP

The NA48 experiment, at the CERN SPS, is a fixed target experiment originally designed to measure direct CPV in neutral kaon decays using simultaneous K$_S$ and K$_L$ beams. After 5 years of data taking NA48 obtained the result $Re(ε'/ε) = (14.7 ±  2.2)·10^{-4}$ [1], which differs from zero by more than six standard deviations. In 2003, the NA48 beam line underwent major modifications in order to accommodate high intensity charged kaon beams.

The NA48 detector [2] consists of two major components: a magnetic spectrometer used as a tracking system for charged particles and a quasi-homogeneous, high granularity liquid Krypton calorimeter (LKr) for the measurement of electromagnetic showers. Both elements allow high resolution detection with very high efficiency. Additional elements, like the charged hodoscope, the hadronic calorimeter, the muon veto system and the large angle photon vetoes, provide extra time information, particle identification and background rejection.

## 2.  CP VIOLATION PARAMETER | η+- |

The parameter η+−  = A(K$_L$→π$^+$π$^-$ )/A(K$_S$→π$^+$π$^-$ ) is a fundamental observable of CPV. It involves both indirect (ε) and direct (ε') modes of CP non-conservation and its magnitude is about 2·10$^{-3}$. Recent precise measurements of η+− by the KTeV [3] and Kloe [4] experiments, give results not in good agreement with previously published values. A precise determination of η+− is therefore important to clarify the experimental situation.  The method used by the NA48 experiment consists in measuring precisely the ratio R = BR(K$_L$→π$^+$π$^-$) / BR(K$_L$→πeν). These decays are characterized by similar signatures involving two reconstructed tracks of charged particles. Then η+− is computed as |η+−|=√ [BR(K$_L$→π$^+$π$^-$) τ$_{KS}$ / BR(K$_S$→π$^+$π$^-$) τ$_{KL}$] using the best single K$_S$ [5] and K$_L$ [6] lifetime measurements and the normalisations BR(K$_L$→πeν) [7] and BR(K$_S$→π$^+$π$^-$) [8] obtained independently by KLOE, KTeV and NA48 the normalisations BR($K_L \to \pi e \nu$) [7] and BR($K_S \to \pi^+\pi^-$) [8] obtained independently by KLOE, KTeV and NA48 experiments.

The analysis is based on the data taken in a 2-day dedicated run in 1999. About $8 \cdot 10^7$ 2-track events were recorded. This statistics allowed the selection of 47,000 reconstructed $K_L \to \pi^+\pi^-$ with a small background contribution of 0.5% originating mainly from the dominant $K_L \to \pi^+\pi^-\pi^0$ and $K_L$ semileptonic decays.

As far as the normalization channel $K_L \to \pi e \nu$ (Ke3) is concerned, the event selection relied on the very good $e/\pi$ separation obtained by comparing the cluster energy E deposited in the LKr calorimeter with the track momentum p measured in the magnetic spectrometer (E/p). About $5 \cdot 10^6$ good Ke3 events were reconstructed with a background level of about 0.5%. The most relevant systematic uncertainties come from the precision of simulation of the kaon energy spectrum, radiative corrections and trigger efficiency measurement.

The ratio R was measured to be R = $(4.835 \pm 0.022_{stat} \pm 0.016_{syst}) \cdot 10^{-3}$ yielding the value $(1.941 \pm 0.019) \cdot 10^{-3}$ for the $K_L \to \pi^+\pi^-$ branching ratio. This result takes into account radiative corrections in which the inner bremsstrahlung component (IB) is included but the CP conserving direct emission process (DE) is subtracted.

Finally, using the experimental inputs described above, the computed value for the | $\eta$+- | parameter was found to be |$\eta$+- | = $(2.223 \pm 0.012) \cdot 10^{-3}$, in agreement with the recent Kloe and KTeV measurements but in contradiction with the published values in PDG2004 [9].

## 3. DIRECT CP VIOLATION PARAMETER $A_g$ IN $K^\pm \to 3\pi$ DECAYS

Due to their high branching fractions, at the level of a few percent, and a simple event selection with low background, $K^\pm \to 3\pi$ decays are promising processes to search for CPV phenomena. NA48/2 used simultaneous $K^+/K^-$ beams and compared the Dalitz plot shapes between $K^+$ and $K^-$ decays into $3\pi$. The $K^\pm \to 3\pi$ decays are usually described in terms of two Lorentz invariant variables u and v, where u=$(s_3-s_0)/m^2_\pi$ and v= $(s_2-s_1)/m^2_\pi$, $s_i = (p_K-p_{\pi i})^2$, $p_K$ is the four momentum of the decaying kaon, and $s_0 = \sum s_i / 3$ (i=1,2,3), with i=3 for the odd pion (the one with different charge from the other two). The $K^\pm \to 3\pi$ matrix element squared is conventionally parametrized as $|M(u,v)|^2 \alpha 1+gu+hu^2+kv^2$ where g, h and k are the so-called linear and quadratic Dalitz plot parameters (|h|, |k| « |g|). A non zero difference $\Delta g$ between the slope parameter $g^+$ and $g^-$ describing the decays of $K^+$ and $K^-$, respectively, is a manifestation of direct CPV expressed by the corresponding slope asymmetry parameter $A_g=(g^+-g^-)/ (g^++g^-) \approx \Delta g/(2g)$. A recent full next-to-leading order ChPT computation [10] predicts $A_g$ to be of the order of $10^{-5}$. Calculations involving processes beyond the SM [11] allow a wider range of $A_g$, including substantial enhancement up to a few $10^{-4}$. The goal of NA48/2 was to measure $A_g$ with an accuracy of a few $10^{-4}$.

The experimental method is based on intense and simultaneous $K^\pm$ beams, superimposed in space, with narrow momentum spectra ($P_K$=60±3 GeV/c). To keep systematic uncertainties at the required level, average $K^+$ and $K^-$ acceptances are equalized by frequently alternating the magnet polarities of the kaon beam lines and of the spectrometer. To extract the slope difference $\Delta g$ we compare the u-projection of the Dalitz plot for $K^+$ and $K^-$ decays:

$$R(u) = N^+(u)/ N^-(u) = 1+(\Delta g \cdot u)/(1+gu+hu^2).$$

However, due mainly to the presence of the magnetic fields, there are experimental asymmetries which do not cancel in this simple ratio. Because of the polarity reversal during data taking, four R(u) ratios with the four possible combinations of magnetic field polarities are defined, the product of these u-ratios forming the quadruple ratio $R_4(u)$. Finally $\Delta g$ is extracted by fitting the quadruple ratio with a function $f(u) = n [1+(\Delta g u)/(1+gu+hu^2)]^4$. This method leads to a cancellation of global time instabilities, local beam line biases and left-right asymmetries. Furthermore it is independent of the $K^+/K^-$ flux ratio, and the analysis does not rely on a detailed Montecarlo simulation.

The result stability is checked with respect to several variables such as kaon energy and decay position, without finding any significant dependence. As a systematic check, "null" ratios are computed by building ratio of events of the same charge, deflected in opposite directions in the spectrometer magnet or distinguished only by the upper or lower path of kaons along the beam line. In this case the result is expected to be equal to zero and any asymmetry in such ratio reflects instrumental biases coupled to time variations. Such effects are at the $10^{-4}$ level, therefore second order effects are negligible, and they will cancel safely in $\Delta g$. Moreover these results are fully reproduced by the Montecarlo simulation as due to time variation of the detector inefficiencies and beam optics.

The selected samples contain $3.11 \cdot 10^9$ $K^\pm \to \pi^+ \pi^+ \pi^-$, and $9.13 \cdot 10^7$ $K^\pm \to \pi^\pm \pi^0 \pi^0$ candidates, practically background free. The CP violating charge asymmetries of the linear slope parameter of the Dalitz plot were respectively found to be:

Charged mode:     $A_g^c = (-1.5 \pm 1.5_{stat} \pm 1.6_{syst}) \cdot 10^{-4} = (-1.5 \pm 2.2) \cdot 10^{-4}$

Neutral mode:     $A_g^0 = (1.8 \pm 1.7_{stat} \pm 0.6_{syst}) \cdot 10^{-4} = (1.8 \pm 1.8) \cdot 10^{-4}$

These results are ten times more precise than any previous measurements. In both modes the statistical errors dominate. The results are compatible with the SM predictions, no evidence for direct CPV is found.

## 4. $V_{US}$ MEASUREMENT FROM $K^\pm_{l3}$

The flavor structure in the quark sector of the SM is described by the CKM matrix. Its unitarity leads to a number of relations for its elements and in particular we have $|V_{ud}|^2+|V_{us}|^2+|V_{ub}|^2 = 1$. Since the contribution for the last term can be neglected this approximation gives $|V_{us}|^2 = \cos \theta_c$ as originally suggested by Cabibbo. The most precise value of $|V_{ud}|$ comes from the super-allowed $0^+ \to 0^+$ beta transitions between nuclei and $|V_{us}|$ is usually calculated from the branching ratios of the semileptonic kaon decays [12].

Going back to PDG2004 [9] $|V_{us}| = 0.2195 \pm 0.0025$ and $|V_{ud}| = 0.9738 \pm 0.0005$ giving a deviation from unitarity at the level of $2.3\sigma$ where the contribution from the uncertainties of $|V_{us}|$ and $|V_{ud}|$ in the final error are almost equal. In the last few years a significant progress in the kaon physics has been made by three experiments Kloe, KTeV and NA48.

More recently NA48/2 has measured the branching ratios of $K^\pm \to \pi^0 e^\pm \nu$ ($K^\pm_{e3}$) and $K^\pm \to \pi^0 \mu^\pm \nu$ ($K^\pm_{\mu3}$) semileptonic decays studying data collected in a dedicated low intensity minimum bias run. Due to the difficulty in measuring the absolute kaon flux we use as normalization channel the decays $K^\pm \to \pi^\pm \pi^0$ ($K^\pm_{2\pi}$). These three modes have very similar topologies and selection criteria and this allows a first order cancellation of systematic effects The basic selection of the three modes is based on the presence of only one charged track in the spectrometer and at least two photon clusters in the LKr consistent with a $\pi^0$ decay. Further kinematical and particle identification requirements are applied to separate the three decays. The invariant mass for $K^\pm_{2\pi}$ candidates is required to be within $3\sigma$ of the reconstructed kaon mass, while for $K^\pm_{l3}$ candidates it has to be outside this range. In order to separate two from three body decays, cuts are applied to the squared missing mass ($m^2_\nu$) [Figure1], using the average energy (60GeV) and direction of the kaon. To distinguish electrons from pions a cut is imposed on the ratio E/p. The identification of muons is based on the associa-tion of a signal in the muon veto detector. The selected samples are practically background free and amount to 87,000 $K^\pm_{e3}$, 77,000 $K^\pm_{\mu3}$ and 718,000 $K^\pm_{2\pi}$ events. After correcting for acceptance, radiative effects, trigger and particle-id efficiencies, we obtain: $R(K^\pm_{e3}/K^\pm_{2\pi}) = 0.2470 \pm 0.0009_{stat} \pm 0.0004_{sys}$ and $R(K^\pm_{\mu3}/K^\pm_{2\pi}) = 0.1636 \pm 0.0006_{stat} \pm 0.0003_{sys}$. Using the new Kloe measurement of BR($K^\pm_{2\pi(\gamma)}$)=0.2065(5)(8) we compute: BR($K^\pm_{e3}$) = 0.05104 $\pm$ 0.00019$_{stat}$ $\pm$ 0.00008$_{sys}$ and BR($K^\pm_{\mu3}$) = 0.03380 $\pm$ 0.00013$_{stat}$ $\pm$ 0.00006$_{sys}$.

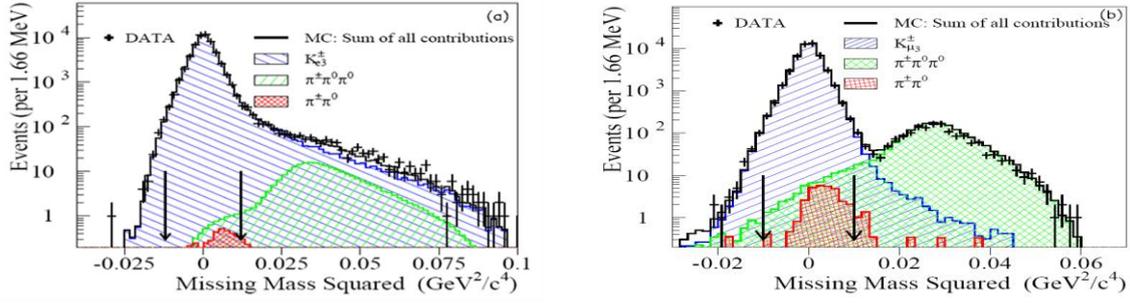

Figure 1: Distribution of squared missing mass ($m^2_\nu$) for $K^\pm_{e3}$ (left) and $K^\pm_{\mu3}$ (right) events. Data vs. Montecarlo.

Using $\tau_{K\pm}$ = 12.370(19)ns (average PDG2006 and Kloe2008) we obtain $|V_{us}| \cdot f_+(0)$ equal to (0.21794 (43)$_{exp}$ (52)$_{norm, \tau}$ (61)$_{ext}$ )=0.2179(9) for $K^\pm_{e3}$, and equal to (0.21818 (46)$_{exp}$ (52)$_{norm,\tau}$ (66)$_{ext}$)=0.2182(10) for $K^\pm_{\mu3}$.

Our two results are in very good agreement and we can combine them taking into account correlation and assuming μ-e universality: $|V_{us}|f_+(0) = 0.2180 \pm 0.0008$, and finally using $f_+(0)$=0.964 $\pm$ 0.005 [RBC-UKQCD2007] we obtain $|V_{us}|$ = 0.2261 $\pm$ 0.0014 in agreement with CKM unitarity when combined with PDG2006 value for $|V_{ud}|$.

Starting from our branching ratios we can also test lepton universality by computing $R(K^\pm_{\mu3}/ K^\pm_{e3})$ = 0.663 $\pm$ 0.003$_{stat}$ $\pm$ 0.001$_{sys}$. This is the most precise single measurement so far and is well consistent with lepton universality which predicts 0.661(3).